\documentclass[12pt,reqno]{amsart}
\usepackage{amssymb,epsf}
\usepackage{amstext,amsmath,amsfonts,amscd,amsthm,graphicx}
\usepackage{upref}
\usepackage{epsfig}
\usepackage[utf8]{inputenc}
\usepackage{latexsym}
\usepackage{eucal}
\usepackage{wrapfig}
\newfont{\fnt}{cmsy10}
\newfont{\sss}{cmti10}

\theoremstyle{definition}

\theoremstyle{plain}
\newtheorem{thm}{Theorem}[section]
\newtheorem{prp}{Proposition}[section]
\theoremstyle{definition}
\newtheorem{rmk}{Remark}[section]

\begin{document}
\author{Barbora Voln\'{a}}
\address{Mathematical Institute, Silesian University in Opava, \newline \indent Na Rybn\'{i}\v{c}ku 1, 746 01 Opava, Czech Republic}
\email{Barbora.Volna@math.slu.cz}

\title{Can we explain unexpected fluctuations of long-term real interest rate?}

\begin{abstract}
In this paper, we present own point of view how the unexpected fluctuations of the long-term real interest rate can be explained. We describe a macroeconomic environment by the modification of the fundamental macroeconomic equilibrium model called the IS-LM model. Last but not least, we suggest a possible cooperation between the fiscal and monetary policy to reduce these fluctuations. Our modelling is demonstrated on an illustrative example.    
\end{abstract}

\maketitle

\section*{Introduction}

In these days many unexpected fluctuations of different economic variables are real and actual and we can observe such situations very often. The mainstream economics tries to explain these fluctuations and these explanations are frequently wrong or debatable. The economy is living and very complex mechanism. Many existing economic theories are imperfect and surely all economic theories depend on their assumptions and this is often overlooked. What I want to say is that it is necessary to look for new economic theories and approaches. That is what I do. My theory do not fit into the existing theories but \linebreak I try to find some new possibilities of the explanation and of the economic interpretation. In this study I focus on unexpected fluctuations of the long-term real interest rate. I try to explain arising unexpected fluctuations and show that these fluctuations can be only seemingly unexpected. E.g. in \cite{cagan}, \cite{henriksen_kydland_sustek}, \cite{holland_toma}, \cite{raunig_scharler} authors studied some fluctuations of interest rates and explained them by various reasons. Even authors in \cite{roley_sellon} pointed out that the connection between monetary policy actions and long-term rates is not reliable. So, I am convinced that focusing to explain some behaviour of the long-term real interest rate is very meaningful.

For the basic description of the economy, we use the fundamental macroeconomic \linebreak IS-LM model. This model explains an aggregate macroeconomic equilibrium, i.e. the goods market equilibrium and the money market (or financial assets market) equilibrium simultaneously. This aged model has been very often modified by many experts, see e.g. \cite{chiba_leong}, \cite{cesare_sportelli}, \cite{gandolfo}, \cite{king}, \cite{neri_venturi} or \cite{zhou_li}. We also create our modification of the original model and we change two original assumptions on more general assumptions. Firstly, we change the constant price level on the floating price level where we are inspired by the IS-ALM model with expectations and the term structure of interest rates, see \cite{baily_friedman}. Secondly, we change the assumption of an exogenous money supply on the assumption of a conjunction of an exogenous and an endogenous money supply. Then we make our modelling based on this modified model. The real presumption of our modelling is that the adjustment speed of the money market is faster than the adjustment speed of the goods market. And the last our assumption relates to the behaviour of the money demand and money supply which are two main quantities of the IS-LM model describing the money market. Based on the mentioned assumptions we show that the special type of the cycle called the relaxation oscillation \cite{bogoliubov_mitropolsky}, \cite{guckenheimer_holmes} exists on the money market in this model. A specific part of the relaxation cycle represents the quick jump or fall of the long-term real interest rate and this is the core of our modelling. This part can be mistakenly interpreted like unexpected. We present the model with once S-bent LM curve, with twice S-bent LM curve and more times S-bent LM curve. The real economies are caused by a fiscal and monetary policy. We show the impacts of these policies on our models. The fiscal policy can cause the relaxation oscillations in the model, i.e. the quick jumps or falls of the long-term real interest rate, and the monetary policy can reduce this negative impact of the fiscal policy. So finally, we suggest the possible cooperation between these two policies. Last section deals with an illustrative example where we show the particular model fulfilling all required conditions and assumptions and the particular suggestion of the fiscal and monetary policy cooperation.

My attitude seems to be little bit controversial because I established the new point of view which may not be consistent with the mainstream economic posture. But according to my opinion it is very important to do something in a different way and to look for another points of view.

\newpage
\section{Description of macroeconomic environment}

In this paper we try to explain a macroeconomic phenomenon, so we start with a description of the studied macroeconomic environment. We deal with an aggregate macroeconomic equilibrium (or stability), i.e. an equilibrium on the goods market and on the money market simultaneously. This is commonly described by the so-called IS-LM model. Firstly, we present the original IS-LM model, and then our modification of this model.

The main models variables are an aggregate income (GDP, GNP) $Y$ and an interest rate $R$. The main models quantities are investments $I$ and savings $S$ describing the goods market, and then a demand for money $L$ and a supply of money $M$ describing the money market.

\subsection{Original IS-LM model}

The original IS-LM model describes a two-sectors economy (the sector of households and the sector of firms); assumes $Y \geq 0,~R>0 $ and \linebreak a constant price level and assumes that the supply is fully adapted to the demand (i.e. a demand-oriented model) and that the money supply is an exogenous quantity (i.e. the money supply is the money stock determined by the central bank), thus $M:=M_S>0$.
      
According to e.g. \cite{gandolfo} the original dynamic IS-LM model is the system of two differential equations
\begin{equation}
\begin{array}{ll}
\textrm{IS:} & \frac{d Y}{d t} = \alpha [I(Y,R)-S(Y,R)] \\
\textrm{LM:} & \frac{d R}{d t} = \beta [L(Y,R)-M_S]
\end{array},
\end{equation}
where $\alpha, \beta >0 $. The IS curve is a set of ordered pairs $[Y,R]$ which satisfy the equation $\alpha[I(Y,R)-S(Y,R)]=0$, i.e. a curve representing the goods market equilibrium. The LM curve is a set of ordered pairs $[Y,R]$ which satisfy the equation $[L(Y,R)-M_S]=0$, i.e. a curve representing the money market equilibrium. An intersection point of the \linebreak IS curve and of the LM curve represents an aggregate macroeconomic equilibrium.

We suppose that all of the considered functions hereinafter are differentiable functions. According to e.g. \cite{gandolfo} the economic properties of the functions $I(Y,R)$, $S(Y,R)$ and $L(Y,R)$ are given by
     \begin{equation}
     \label{economic_I} 
     0<\frac{\partial I}{\partial Y}<1, \frac{\partial I}{\partial R}<0,
     \end{equation}
     \begin{equation}
     \label{economic_S}
     0<\frac{\partial S}{\partial Y}<1, \frac{\partial S}{\partial R}>0,
     \end{equation}
     \begin{equation}
     \label{economic_L_Y}
     \frac{\partial L}{\partial Y}>0,
     \end{equation}     
     \begin{equation}
     \label{economic_L_R}
     \frac{\partial L}{\partial R}<0. 
     \end{equation}

\subsection{New IS-LM model}

In this subsection, we present a modification of the original \linebreak IS-LM model on which other models are based, and we present the other properties of the main model functions too. For this model modification see also \cite{volna_MME} and \cite{volna_AMC}.

This model modification is named the new IS-LM model. As the original IS-LM model, the new IS-LM model describes a two sectors economy; assumes $Y \geq 0$ and assumes that the supply is fully adapted to the demand. In addition, the new IS-LM model assumes \linebreak a variable price level, thus this model distinguishes two types of the interest rates - the long-term real interest rate $R$, $R \in \mathbb{R}$, and the short-term nominal interest rate $i_S$, $i_S > 0$, and the new IS-LM model assumes a conjunction of the endogenous money supply and of the exogenous money supply, i.e. the money stock is assumed to be generated by the credit creation with the partial central bank determination.

Generally, it is considered that the long-term real interest rate affects the goods market and the short-term nominal interest rate affects the money market. We use the well-known formula, see e.g. \cite{baily_friedman},
\begin{equation}
\label{relation_i_S_R}
i_S = R - MP + \pi^e ,
\end{equation}
where $MP$ is a maturity premium and $\pi^e$ is an expected inflation rate, for express \linebreak a mutual relation of these two types of the interest rates. 

Then we define the money supply as partly endogenous and partly exogenous quantity using the formula
\begin{equation}
  M(Y,i_S) + M_S  ,
\end{equation}
where the function $M(Y,i_S)$ represents the endogenous part of the money supply (i.e. the part of the money stock generated in the economy by the credit creation, see e.g. \cite{sojka}) and the constant $M_S>0$ represents the exogenous part of the money supply (i.e. the part of the money stock determined by the central bank).

\begin{rmk}
As many empirical studies show, see e.g. \cite{badarudin_ariff_khalid_1}, \cite{badarudin_ariff_khalid_2}, \cite{badarudin_khalid_ariff}, \cite{howells}, the endogenous and also exogenous money supplies exist in many economies all of the World. Also some theoretical work describes the existence of both types of the money supply, see e.g. \cite{sedlacek}.
\end{rmk}

For simplification we assume constant $MP$ and $\pi^e$, then $\frac{d i_S}{d t} = \frac{d (R - MP + \pi^e)}{d t} = \frac{d R}{d t}$ and $\frac{\partial L(Y,i_S)}{\partial i_S}=\frac{\partial L(Y,R - MP + \pi^e)}{\partial R}$, $\frac{\partial M(Y,i_S)}{\partial i_S}=\frac{\partial M(Y,R - MP + \pi^e)}{\partial R}$. We establish the properties of the money supply function as it is described below:
\begin{equation}
\label{economic_M_Y} 
0 < \frac{\partial M}{\partial Y} < \frac{\partial L}{\partial Y},
\end{equation}
\begin{equation}
\label{economic_M_R} 
\frac{\partial M}{\partial i_S}=\frac{\partial M}{\partial R}>0.
\end{equation}
The condition (\ref{economic_M_Y}) includes also the property (\ref{economic_L_Y}). 

\begin{rmk}
\label{rmk_M_Y_and_M_R}
Let us try to examine the properties (\ref{economic_M_Y}) and (\ref{economic_M_R}). The positive relation between the money supply and the aggregate income follows from the reverse causality theory based on the expectations of the subjects on the money market and their reactions. E.g. in \cite{almonacid} the positive relation between these two variables is showed but the causality direction from the money supply to the income is showed. E.g. in \cite{bozoklu} based on empirical tests the causality direction from the income to the money supply is showed. Then we assume that the rate of increase of the money supply depending on the aggregate income is smaller than the rate of increase of the money demand depending on the aggregate income. Our explanation of such condition is that the banks (which indirectly produce money by the credit creation) in their expectations are more cautious than other subjects on the money market. But we do not need identify with this money supply property. If we do not compare the rate of increase between the money demand and the money supply then we can add (\ref{economic_M_Y}) to our modelling as a necessary condition. Now, we are focusing on the property (\ref{economic_M_R}). If we consider that the interest rate is a "price" of the money and there is causality from the interest rate to the money supply, we can see that the "classical" principle of supply may hold (with the difference that we are on the money market). So, the higher price (interest rate) implies the higher supply (of money). Suppliers are commercial banks through the offers of loans. This view is also more less consistent with the theory of the relative money endogeneity, see e.g. \cite{rousseas}. Also some empirical evidence suggests the positive relationship between the interest rate and the money supply, see e.g. \nolinebreak \cite{monnet_weber}. 
\end{rmk}

Summary, compared to the original IS-LM model the LM equation is changed. Now, the money demand is described by the function $L(Y,i_S)=L(Y,R - MP + \pi^e)$ and the money supply is described by the formula  $M(Y,i_S)+M_S=M(Y,R - MP + \pi^e)+M_S$. The \textit{new dynamic IS-LM model} is given by the system of two differential equations
\begin{equation}
\label{new_dynamic_IS-LM}
\begin{array}{ll}
\textrm{IS:} & \frac{d Y}{d t} = \alpha [I(Y,R)-S(Y,R)] \\
\textrm{LM:} & \frac{d R}{d t} = \beta [L(Y,R- MP + \pi^e)-M(Y,R - MP + \pi^e)-M_S]
\end{array},
\end{equation}
where $\alpha, \beta >0 $.

In addition to the main model functions properties (\ref{economic_I}), (\ref{economic_S}), (\ref{economic_L_R}), (\ref{economic_M_Y}) and (\ref{economic_M_R}) we assume 
\begin{equation}
\label{economic_I_S}
\frac{\partial I}{\partial Y} < \frac{\partial S}{\partial Y},
\end{equation}
which represents the most typical economic case.

\section{Models of seemingly unexpected fluctuations \\ of long-term real interest rate}

The starting point of the creation of these models is the new IS-LM model presented in the previous section. Moreover, we assume two additional conditions and we try to give some economic interpretation of these conditions. Firstly, we assume the real property of model markets that the adjustment speed of the money market is greater than the adjustment speed of the goods market. The second condition relates to a specific behaviour of the money demand and of the money supply leading to the particular shape of the LM curve. The first model is the model with once S-bent LM curve. The second model is the model with twice S-bent LM curve. The last subsection briefly outlines the problems of the models with more times S-bent LM curve. In these models so-called relaxation oscillations emerge on the LM side of the model, i.e. on the money market. This special type of the oscillation contains parts where the long-term real interest rate quickly jumps or falls and these parts can be mistakenly interpreted like unexpected fluctuations of the long-term real interest rate.

\subsection{Model with once S-bent LM curve}

In this subsection, we formulate the model with once S-bent LM curve. As mentioned above, we assume the real presumption that the adjustment speed of the money market is greater than the adjustment speed of the goods market, i.e. the subjects on the money market are faster in their reactions than the subjects on the goods market. The time change of the long-term real interest rate \nolinebreak $R$ describes the money market velocity and the time change of the aggregate income $Y$ describes the goods market velocity. So, the situation where $Y$ is changing very slowly in time in proportion to $R$ can be described by the system, see \cite{volna_MME},
\begin{equation}
\label{dynamic_model_LM_parameter} 
  \begin{array}{lll}
  \frac{d Y}{d t} & = & \varepsilon \alpha [I(Y,R)-S(Y,R)]  \\
  \frac{d R}{d t} & = & \beta [L(Y,R - MP + \pi^e)-M(Y,R - MP + \pi^e)-M_S]  
  \end{array},
\end{equation}
where $0 < \varepsilon << 1$ is a very small positive parameter. Therefore, we can consider $\frac{d Y}{d t} = 0$ because of very small $\varepsilon$ and we can rewrite the system (\ref{dynamic_model_LM_parameter}) to the system, see \cite{volna_MME}, 
\begin{equation}
\label{dynamic_model_LM_parameter_2} 
  \begin{array}{lll}
  \frac{d Y}{d t} & = & 0  \\
  \frac{d R}{d t} & = & \beta [L(Y,R-MP+\pi^e)-M(Y,R-MP+\pi^e)-M_S] 
  \end{array}.
\end{equation}
In the system (\ref{dynamic_model_LM_parameter_2}) only the LM curve remains to be dealt with and $Y$ becomes a parameter in the equation $\frac{d R}{d t}=\beta[L(Y,R-MP+\pi^e)-M(Y,R-MP+\pi^e)-M_S]$.

\begin{figure}[ht]
  \centering
  \includegraphics[height=5cm]{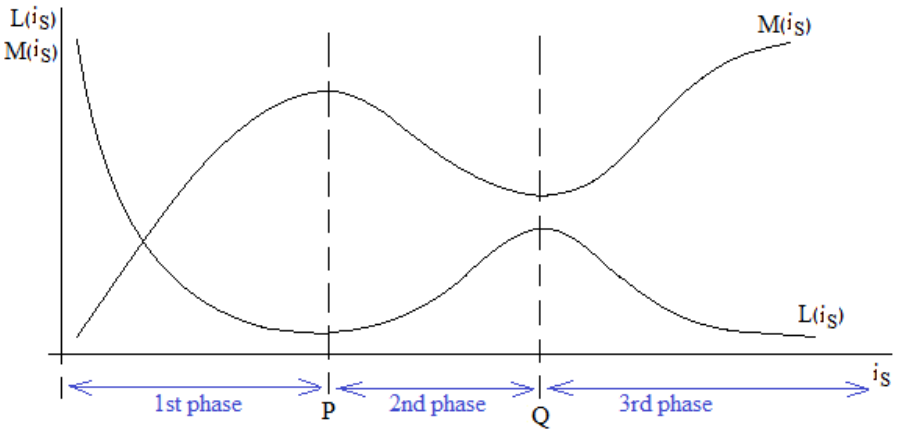}
  \caption{Three phases of $L(i_S)$ and $M(i_S)$, see \cite{volna_MME}}
  \label{fig:L(i_S)_and_M(i_S)_3_phases}
\end{figure}

Now, we present the additional condition of the particular money demand and money supply behaviour as mentioned above. We call this condition the \textit{three phases money demand and money supply} depending on the short-term nominal interest rate $i_S$ for some fixed aggregate income $Y$, see also \cite{volna_MME}. As the name of this condition suggest both functions (of the money demand $L(i_S)$ and of the money supply $M(i_S)$) have three phases of their courses: the first phase for $i_S \in [0, P), P>0$, the second phase for $i_S \in (P, Q), P < Q$ and the third phase for $i_S \in (Q, \infty)$. In the first phase and in the third phase, the money market subjects behave usually, i.e. $L(i_S)$ and $M(i_S)$ have properties (\ref{economic_L_R}) and (\ref{economic_M_R}). In the second phase, the behaviour of the money market subjects is precisely reversed, i.e $L(i_S)$ and $M(i_S)$ have properties
\begin{equation}
\label{unusual_economic_L_M_R} 
\frac{\partial L}{\partial i_S}=\frac{\partial L}{\partial R} >0,~\frac{\partial M}{\partial i_S}=\frac{\partial M}{\partial R} <0.
\end{equation}
And $\frac{\partial L}{\partial i_S}=\frac{\partial M}{\partial i_S}=0$ for $i_S=P$ and for $i_S=Q$. On Figure \ref{fig:L(i_S)_and_M(i_S)_3_phases}, the graphs of the three phases money demand and money supply function (depending on $i_S$ for some fixed $Y$) are displayed.

\begin{rmk}
\label{rmk:three_phases_M_L}
Let us try to examine the three phases money demand and money supply from the economic point of view. The first and the third phase correspond to the usual behaviour of the money demand and supply as it is described also in Remark \ref{rmk_M_Y_and_M_R}. The second phase is different. If we consider that the money market principle resembles the goods market principle, i.e. the interest rate is considered to be a "price" of the money and there exist the supply and demand principles, we can understand this phase like \linebreak a similarity to the Giffen good behaviour. So, if the interest rate increases, the demand for money rises. We assume the demand-oriented model, then the supply fully adapts to the demand, and so if the interest rate rises, the money supply decreases. We can find that this behaviour can be similar to the liquidity trap behaviour joined with the so-called Keynes effect. This situation corresponds to pessimistic expectations of the economic subjects. It is necessary to remark that reasons and repercussions of the liquidity trap differ under the various theories. E.g. in the well-known real case of the liquidity trap in Japan in the nineties the recommended fiscal solutions did not improve the situation. According to some empirical studies, e.g. \cite{hnatkovska_lahiri_vegh} or \cite{monnet_weber}, the reverse behaviour of the money demand and of the money supply function described by (\ref{unusual_economic_L_M_R}) is admitted. The last point is an explanation of the turn from the original behaviour to the precisely reversed behaviour of the money demand and supply, and vice versa. According to the theory of the relative money endogeneity there exists some maximal level of an interest rate joined with the maximal risk level of banks to provide loans, so there exists also the maximal level of the money supply generated by the credit creation. This situation can lead to the pessimistic expectations of the economic subjects and thus to the change their behaviour. This change corresponds to the point $P$ in graphs on Figure \ref{fig:L(i_S)_and_M(i_S)_3_phases}. The contrary change of the economic subjects behaviour corresponding to the point $Q$ can be explained like a "saturation point". In this point the economic subjects change their expectations to more optimistic expectations and the subjects representing the demand side are satisfied by holding an amount of money. Then with a raising interest rate they more prefer the holding of stocks than the holding of liquid assets and the money demand decreases. The money supply adapts to the money demand and so the money supply increases with the raising interest rate as in the first phase.
\end{rmk}

\begin{figure}[ht]
  \centering
  \includegraphics[height=4.5cm]{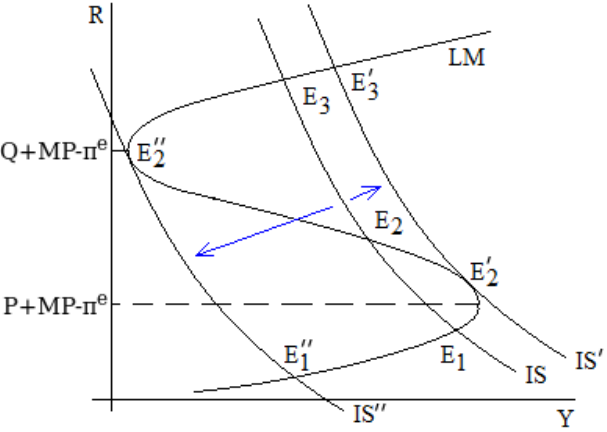}
  \caption{Shifting IS curve and LM curve in this case}
  \label{fig:IS_LM_curves}
\end{figure}

The IS curve and the LM curve in the new IS-LM model (\ref{new_dynamic_IS-LM}) or rather in the system (\ref{dynamic_model_LM_parameter}) with properties (\ref{economic_I}), (\ref{economic_S}), (\ref{economic_M_Y}), (\ref{economic_I_S}) and with three phases money demand and money supply are depicted on Figure \ref{fig:IS_LM_curves}. The slope of the LM curve is given by $-\frac{L_Y-M_Y}{L_R-M_R}$ and the slope of the IS curve is given by $-\frac{I_Y-S_Y}{I_R-S_R}$ according to Implicit Function Theorem. From the condition (\ref{economic_M_Y}) and three phases money demand and money supply it follows that the LM curve is increasing for $i_S \in (-\infty,P) \cup (Q,\infty)$, i.e. for $R \in (-\infty,P+MP-\pi^e) \cup (Q+MP-\pi^e,\infty)$, and decreasing for $i_S \in (P,Q)$, i.e. for $R \in (P+MP-\pi^e,Q+MP-\pi^e)$. From the conditions (\ref{economic_I}), (\ref{economic_S}) and (\ref{economic_I_S}) it follows that IS curve is decreasing for all $R \in \mathbb{R}$. In such a system, there can occur at least one and the most three singular points, see Figure \ref{fig:IS_LM_curves} and shift the IS curve downwards or upwards.

\begin{figure}[ht]
  \centering
  \includegraphics[height=4.5cm]{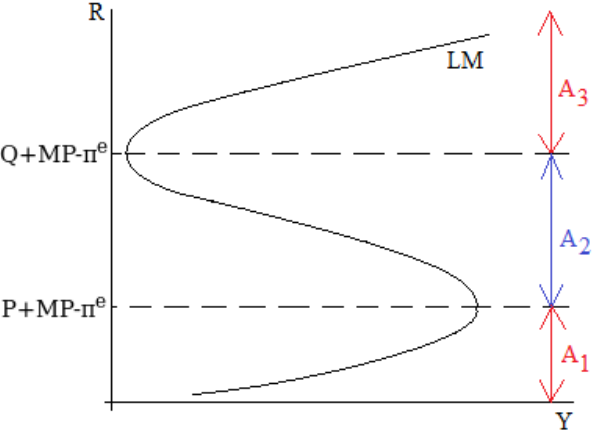}
  \caption{Arcs of the LM curve}
  \label{fig:LM_curve}
\end{figure}

Now, we consider the situation where the adjustment speed of the money market is greater than the adjustment speed of the goods market described by the system (\ref{dynamic_model_LM_parameter_2}). The LM curve is the curve defined by the equation $\beta [L(Y,R - MP + \pi^e)-M(Y,R - MP + \pi^e)-M_S]=0$ and the LM curve is displayed on Figure \ref{fig:LM_curve}. Every point of the LM curve corresponding to this system is singular point. We need to distinguish the stable and unstable arcs of the LM curve, see Proposition \ref{prp:stable_unstable_arcs}.

\begin{prp}
\label{prp:stable_unstable_arcs}
The arc $A_1$ and the arc $A_2$ of the LM curve are stable arcs and the arc $A_2$ of the LM curve is unstable arc.
\end{prp}

\begin{proof}
The qualification of the singular points of the new IS-LM model (\ref{dynamic_model_LM_parameter}) with conditions (\ref{economic_I}), (\ref{economic_S}), (\ref{economic_M_Y}), (\ref{economic_I_S}) and with three phases money demand and money supply for $i_S$ determines the stability and unstability of the LM curve arcs in the system (\ref{dynamic_model_LM_parameter_2}). At least one and at the most three singular points can exist in this system, see Figure \ref{fig:IS_LM_curves}. Let $J$ denote the Jacobi's matrix of the system (\ref{dynamic_model_LM_parameter}) and $\lambda_{1,2}$ denote the eigenvalues of this Jacobi's matrix. $ \lambda_{1,2} = \frac{1}{2} \left[ \varepsilon \alpha (I_Y - S_Y) + \beta (L_R-M_R) \pm \sqrt{\left[ \varepsilon  \alpha (I_Y - S_Y) + \beta (L_R-M_R) \right]^2 - 4detJ} \right]$ where $det J = \varepsilon  \alpha \beta \left[ (I_Y - S_Y)(L_R-M_R) - (I_R - S_R)(L_Y-M_Y) \right]$. Firstly, we consider the singular points located on the arc $A_1$ or $A_3$ (see e.g. $E_1$ or $E_3$ on Figure \ref{fig:IS_LM_curves}). From the properties (\ref{economic_I}), (\ref{economic_S}),  (\ref{economic_M_Y}), (\ref{economic_I_S}) and (\ref{economic_L_R}) with (\ref{economic_M_R}) (points are situated in the area of $R \in (-\infty,P+MP-\pi^e] \cup [Q+MP-\pi^e, \infty)$) it follows that $Re(\lambda_{1,2}) < 0$ and $det J>0$ in these points. Thus, these points and so the arc $A_1$ and $A_3$ are stable. Secondly, we consider the singular points located on the arc $A_2$ (see e.g. $E_2$ or $E_2'$ on Figure \ref{fig:IS_LM_curves}). We distinguish two possibilities: with three singular points in the model (see e.g. $E_1$, $E_2$ and $E_3$ on Figure \ref{fig:IS_LM_curves}) and with two singular points in the model (see e.g. $E_2'$ and $E_3'$, or $E_1''$ and $E_2''$ on Figure \ref{fig:IS_LM_curves}). In the first possibility the slope of the IS curve is smaller than the slope of the LM curve in the point located on the arc $A_2$ (see e.g. $E_2$), i.e. $-\frac{I_Y - S_Y}{I_R - S_R}<-\frac{L_Y-M_Y}{L_R-M_R}$. This condition and properties (\ref{economic_I}), (\ref{economic_S}), (\ref{economic_M_Y}), (\ref{economic_I_S}) and (\ref{unusual_economic_L_M_R}) (points are situated in the area of $R \in (P+MP-\pi^e,Q+MP-\pi^e)$) imply $det J<0$ in these points. Thus these points are unstable and so the arc $A_2$ is unstable. In the second possibility the IS curve and the LM curve have the same slope, i.e. $-\frac{I_Y - S_Y}{I_R - S_R}=-\frac{L_Y-M_Y}{L_R-M_R}$. This implies zero determinant of $J$ and at least one zero eigenvalue in these points. These singular points are unstable. Thus, $A_2$ is unstable.
\end{proof}

\begin{thm}
Let us consider a new IS-LM model with very slow changes of aggregate income $Y$ in time (\ref{dynamic_model_LM_parameter}) and with three phases money demand and money supply subjected to the economic conditions (\ref{economic_I}), (\ref{economic_S}), (\ref{economic_M_Y}), (\ref{economic_I_S}). Then in such a model the counterclockwise relaxation oscillations arise.
\end{thm}

\begin{figure}[ht]
  \centering
  \includegraphics[height=5.9cm]{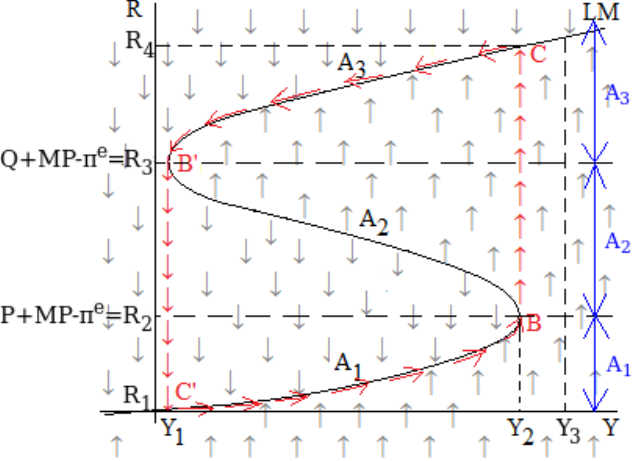}
  \caption{Relaxation oscillations in the model with once S-bent LM curve}
  \label{fig:one_S-bend_rel_osc}
\end{figure}

\begin{proof}
The simplified system (\ref{dynamic_model_LM_parameter_2}) instead the system (\ref{dynamic_model_LM_parameter}) is considered. The three phases money demand and money supply and the economic property (\ref{economic_M_Y}) ensure the form of the LM curve, see Figure \ref{fig:LM_curve}, and every point of the LM curve is stationary point. We can see the vector field given by the right-hand side of the system (\ref{dynamic_model_LM_parameter_2}) on Figure \ref{fig:one_S-bend_rel_osc}. The trajectories of this system have the up or down vertical direction almost everywhere, except the parts corresponding to the close neighbourhood of the stable arcs $A_1$ and $A_3$ where the moving point goes along the LM curve in the direction depicted by red arrows on this figure. The sign of the function $\alpha [I(Y,R)-S(Y,R)]$ on the LM curve determines this vertical direction of the trajectories, i.e. this direction is determined by the stability or unstability of the relevant arcs of the LM curve, see Proposition \ref{prp:stable_unstable_arcs}. We can see that $Y_1, Y_2$ are bifurcations values. Thus, one stable stationary state exists in the area for $Y<Y_1$, $Y>Y_2$ and three stationary states exist in the area for $Y_1<Y<Y_2$: two stable states ($A_1$ and $A_3$) and one unstable state between them ($A_2$). The stable arcs $A_1$ and $A_3$ attract the moving point and the unstable arc $A_2$ repel the moving point. There is an infinite large velocity of the motion almost everywhere, except the parts with finite large velocity of the motion situated very close to the LM curve. Now, we show the construction of the relaxation oscillations. The one counterclockwise cycle, see the red cycle on Figure \ref{fig:one_S-bend_rel_osc}, can be divided by the points $B$, $C$, $B'$, $C'$ to the four parts. The counterclockwise orientation is given by the shape of the LM curve. Let us consider that the parameter $Y$ is changed very slowly from the level $Y_2$ to the level $Y_1$, or from the level $Y_1$ to the level $Y_2$, and the moving point is on or close to the stable arc $A_3$, or on or close to the stable arc $A_1$, respectively. Then the moving point goes along this stable arc until the point $B'$, or until the point $B$, respectively. Then the moving point passes the unstable arc $A_2$ and is attracted to the point $C'$ located on the stable arc $A_1$, or to the point $C$ located on the stable arc $A_3$, respectively. The velocity of the motion between the point $B'$ and $C'$, or $B$ and $C$ is infinitely large. Summary, two vertical segments with infinitely large velocity of the motion (i.e. the part of the red cycle between the point $B$ and $C$ and the part of the red cycle between the point $B'$ and $C'$ on Figure \ref{fig:one_S-bend_rel_osc}) and two segment corresponding to the trajectories along the stable arc $A_1$ and $A_3$ with finite velocity of the motion (i.e. the part of the red cycle between the point $C$ and $B'$ and the part of the red cycle between the point $C'$ and $B$ on Figure \ref{fig:one_S-bend_rel_osc}) form the relaxation oscillations.
\end{proof}

The mentioned vertical segments with the infinitely large velocity of the motion constitute the quick falls or jumps of the long-term real interest rate. The existence of the relaxation oscillations on the money market described by the presented system with given conditions can explain the fluctuations of the long-term real interest rate which can be initially considered as unexpected.

\subsection{Model with twice S-bent LM curve}

\begin{figure}[ht]
  \centering
  \includegraphics[height=6cm]{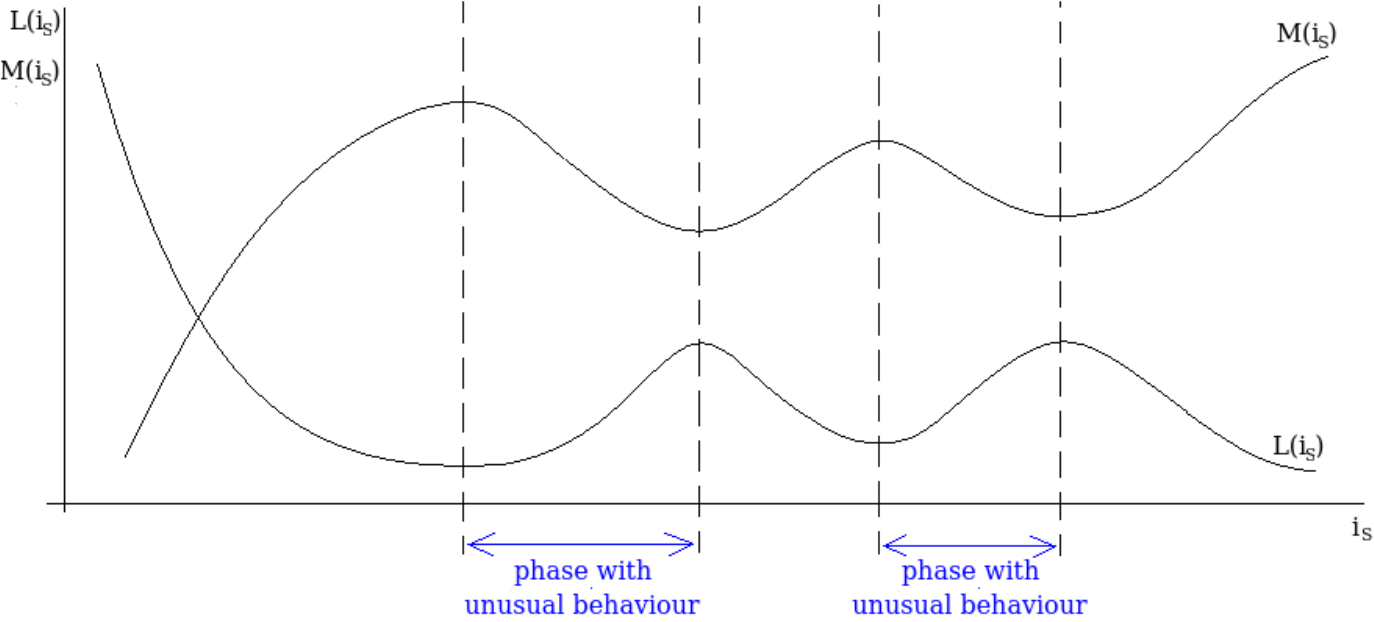}
  \caption{Graphs of $L(i_S)$ and $M(i_S)$ with two phases with unusual behaviour}
  \label{fig:two_S-bends_L(i_S)_and_M(i_S)}
\end{figure}

In this subsection, we briefly present the illustrative model with twice S-bent LM curve. Every specific system with twice S-bent LM curve has qualitatively the same dynamical behaviour. We consider the system (\ref{dynamic_model_LM_parameter}), or rather (\ref{dynamic_model_LM_parameter_2}) and economic properties of the model functions as in the previous section except the condition named the three phases money demand and money supply which is extended. In this situation the bends of the LM curve are ensured by this economic condition of the money demand and of the money supply function depending on the short-term nominal interest rate. If there are two parts with unusual behaviour of these functions described by (\ref{unusual_economic_L_M_R}) between three usual parts described by (\ref{economic_L_R}) and (\ref{economic_M_R}), see Figure \nolinebreak \ref{fig:two_S-bends_L(i_S)_and_M(i_S)}, then there exists a twice S-bent LM curve in such systems according to Implicit Function Theorem. The economic interpretation of the course of the money demand and of the money supply function with three usual parts and two unusual parts is quite similar as in the Remark \ref{rmk:three_phases_M_L}.

\begin{figure}[ht]
  \centering
  \includegraphics[height=10cm]{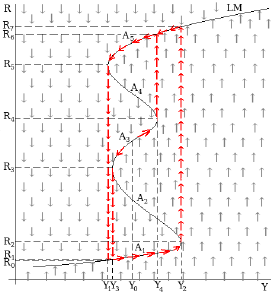}
  \caption{Relaxation oscillations in the model with twice S-bent LM curve}
  \label{fig:two_S-bends_rel_osc}
\end{figure}

We can see the dynamical behaviour in this case, i.e. the relaxation oscillations emerging in these systems with twice S-bent LM curve, on Figure \ref{fig:two_S-bends_rel_osc}.
The arcs $A_1$, $A_3$ and $A_5$ are stable arcs and arcs $A_2$ and $A_4$ are unstable arcs. Similarly as in the previous case, the trajectories have the up or down vertical direction everywhere except the close neighbourhood of the stable arcs $A_1$, $A_3$ and $A_5$ where the trajectories copy these arcs in the orientation depicted by the red arrows in Figure \ref{fig:two_S-bends_rel_osc}. The velocity of the motion along the trajectory is finite close to the LM curve and infinitely large elsewhere given by attracting stable arcs and repelling unstable arcs. The direction of the path and velocity of motion of the moving point are dependent on its current position and on the changes of the parameter \nolinebreak $Y$. If the moving point is close to the stable arc $A_1$ and if we are changing parameter $Y$ from the level $Y_1$ to $Y_2$, then this moving point will go along the stable arc $A_1$ and then the moving point will be attracted from the point $[Y_2,R_2]$ to the point $[Y_2,R_7]$. So, the long-term real interest rate jumps from the level $R_2$ to $R_7$. Then if the moving point is close to the stable arc $A_5$ and if we are changing parameter $Y$ from the level $Y_2$ to $Y_1$, the moving point will go along the stable arc $A_5$ and then the moving point will be attracted from the point $[Y_1,R_5]$ to the point $[Y_1,R_0]$. So, the long-term real interest rate falls from the level $R_5$ to $R_0$. So, we have similar counterclockwise cycle as in the previous model. But if the moving point is close to the stable arc $A_3$ and we are changing parameter $Y$ for example from the level $Y_0$ to the level $Y_3$, the moving point will go along the stable arc $A_3$ and then the moving point will be attracted from the point $[Y_3,R_3]$ to the point $[Y_3,R_1]$. So the long-term real interest rate falls from the level $R_3$ to $R_1$. After that the moving point will join to previously described cycle. And if the moving point is close to the stable arc $A_3$ and we are changing parameter $Y$ for example from the level $Y_0$ to the level $Y_4$, the moving point will go along the stable arc $A_3$ and then the moving point will be attracted from the point $[Y_4,R_4]$ to the point $[Y_4,R_6]$. So the long-term real interest rate jumps from the level $R_4$ to $R_6$. After that the moving point will join to previously described cycle. In this illustrative model, we can observe four seemingly unexpected fluctuations of $R$: from $R_2$ to $R_7$, from $R_5$ to $R_0$, from $R_3$ to $R_1$ and from $R_4$ to $R_6$, see Figure \ref{fig:two_S-bends_rel_osc}.

\subsection{Models with more times S-bent LM curve}

In this subsection, we illustrate the models with more times S-bent LM curve on the one sketch of the vector field given by the right-hand side of the relevant system where three times S-bent LM curve is found. We consider the previously presented new IS-LM model with all relevant assumptions. The number of bends of the LM curve follows from the number of parts describing an unusual behaviour of the money demand and of the money supply depending on the short-term nominal interest rate.

\begin{figure}[ht]
  \centering
  \includegraphics[height=11cm]{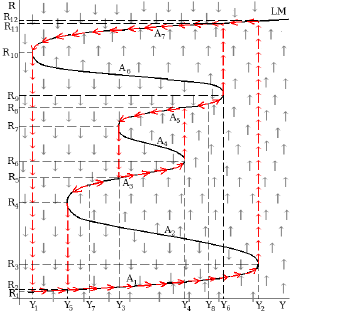}
  \caption{Relaxation oscillations in the model with three times S-bent LM curve}
  \label{fig:three_S-bends_rel_osc}
\end{figure}

We can see the dynamical behaviour of this particular system on Figure \ref{fig:three_S-bends_rel_osc}.
The relaxation oscillations emerge here just as in the previously presented models. $A_1$, $A_3$, $A_5$ and $A_7$ are stable arcs and $A_2$, $A_4$ and $A_6$ are unstable. There exist two counterclockwise cycles. The first cycle consists of two vertical segments with infinitely large velocity of the motion (i.e. the segment from the point $[Y_2,R_3]$ to the point $[Y_2,R_{12}]$ and from $[Y_1,R_{10}]$ to $[Y_1,R_1]$) and of two parts of the stable arcs $A_1$ and $A_7$ with finite velocity of the motion. The second cycle consists of two vertical segments with infinitely large velocity of the motion (i.e. the segment from $[Y_4,R_6]$ to $[Y_4,R_8]$ and the segment from $[Y_3,R_7]$ to $[Y_3,R_5]$) and of two parts of the stable arcs $A_3$ and $A_5$ with finite velocity of the motion. But if the moving point is close to the stable arc $A_3$ and if the parameter $Y$ is increasing from the level $Y_7$ then the moving point will go along the stable arc $A_3$ and then the moving point will join to the second mentioned cycle. Similarly, if the moving point is close to the stable arc $A_3$ and if the parameter $Y$ is decreasing from the level $Y_7$ then the moving point will go along the stable arc $A_3$ and then the moving point will be attracted from the point $[Y_5,R_4]$ to the point $[Y_5,R_2]$ and then the moving point will join to the first mentioned cycle. Analogous situation is for the position of the moving point close to the stable arc $A_5$. Summary, there we can observe six seemingly unexpected fluctuations of $R$: from $R_{10}$ to $R_1$, from $R_3$ to $R_{12}$, from $R_7$ to $R_5$, from $R_6$ to $R_8$, from $R_4$ to $R_2$ and from $R_9$ to $R_{11}$, see Figure \ref{fig:three_S-bends_rel_osc}.

\section{Influences of fiscal and monetary policy}

\subsection{Fiscal policy - changing of aggregate income $Y$}

The fiscal policy can cause the relaxation oscillations on the money market, so the fiscal policy can cause the mentioned quick jumps or falls of the long-term real interest rate. The core of this government influences lies in the possibility to change the level of the aggregate income using its own tools for do this. We demonstrate the impact of a changing of the level of the aggregate income on the model with once S-bent LM curve using Figure \ref{fig:one_S-bend_rel_osc}. If the government decides to increase the aggregate income from the level $Y_1$ to the level $Y_3$, then the long-term real interest rate will be increasing slowly (along the stable arc $A_1$), then for the level of the aggregate income $Y_2$ the long-term real interest rate will jump from the level $R_2$ to the level $R_4$ and then the long-term real interest rate will be increasing very slowly again (along the stable arc $A_3$) until aggregate income will be on the level $Y_3$. There can be similar situation if the government decides to decrease the aggregate income. The long-term real interest rate would decrease very slowly except the point for the level of the aggregate income $Y_1$ where the long-term real interest rate would fall from the level $R_3$ to the level \nolinebreak $R_1$.

In the model with several times S-bent LM curve there can be several jumps or falls of the long-term real interest rate. In our presented model with three times S-bent \linebreak LM curve, see Figure \ref{fig:three_S-bends_rel_osc}, the economy is represented for example by the point on the stable arc $A_3$ for the level of the aggregate income $Y_7$. Then the government decides to increase the aggregate income to the level $Y_6$. Then the long-term real interest rate will be growing slowly along the stable arc $A_3$, for the level of the aggregate income $Y_4$ the long-term real interest rate will jump from the level $R_6$ to the level $R_8$, then the long-term real interest rate will be growing slowly along the stable arc $A_5$ until the aggregate income will be on the level $Y_6$ where the long-term real interest rate will jump from the level $R_9$ to the level $R_{11}$.

\begin{rmk}
The fiscal policy can be also demonstrated by shifting of the IS curve in the new IS-LM model. If the government increases the aggregate income, the IS curve is shifted upwards and vice versa. So, we can show the relaxation oscillations using the both curves by shifting IS curve downwards and upwards. On Figure \ref{fig:IS_LM_curves}, we can see three singular points $E_1$, $E_2$ and $E_3$ (intersection points of the IS curve and of the LM curve). $E_1$ and $E_3$ are stable singular points, $E_2$ is unstable point. Then the IS curve shifts upwards or downwards according to the changing of the aggregate income $Y$ until there are two singular points $E_2'$ and $E_3'$ (intersection points of the IS' curve and of the LM curve), or $E_1''$ and $E_2''$ (intersection points of the IS'' curve and of the LM curve). The points $E_1''$ and $E_3'$ are stable points and the points $E_2'$ and $E_2''$ are unstable points. If the government increases the aggregate income, the IS curve is shifted slowly upwards and the general equilibrium point is shifted slowly along the LM curve for example from the the point $E_1$ to the point $E'_2$. Then the general equilibrium point is shifted quickly from the point $E'_2$ to the point $E'_3$ along the moved IS curve (denoted by IS'). Similarly, if the government decreases the aggregate income, the IS curve is shifted slowly downwards and the general equilibrium point is shifted slowly along the LM curve for example from the the point $E_3$ to the point $E''_2$. Then the general equilibrium point is shifted quickly from the point $E''_2$ to the point $E''_1$ along the moved IS curve (denoted by IS'').
\end{rmk}

\subsection{Monetary policy - changing of expected inflation rate $\pi^e$ or constant part of money stock $M_S$}

The central bank implements a monetary policy. The central bank can only indirectly influences the long-term real interest rate by an inflation targeting. The households and firms adapt an expected inflation rate according to the prediction of the central bank. In accordance with this, we demonstrate the monetary policy by changing of an expected inflation rate $\pi^e$ in our models. A determination of the constant part of the money stock $M_S$ is another tool of the monetary policy to influence the money market. Both of these changes cause the shift of the LM curve up or down.

\begin{figure}[ht]
  \centering
  \includegraphics[height=8.2cm]{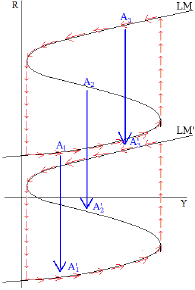}
  \caption{Demonstration of the monetary policy}
  \label{fig:demonstration_monetary_policy}
\end{figure}

If the central bank increases the inflation rate $\pi^e$, then the long-term real interest rate \nolinebreak $R$ decreases because of (\ref{relation_i_S_R}). Then the LM curve is shifted downwards. If the central bank increases the controlled money stock $M_S$, then the LM curve is also shifted downwards because of the LM equation of the model (\ref{new_dynamic_IS-LM}) and so of the model (\ref{dynamic_model_LM_parameter}) and (\ref{dynamic_model_LM_parameter_2}). And vice versa, the LM curve is shifted upwards in the case of the decreasing inflation rate $\pi^e$ or of the decreasing controlled money stock $M_S$. We can illustrate this mechanism on the model with once S-bent LM curve, see Figure \ref{fig:demonstration_monetary_policy}.

\subsection{Cooperation between fiscal and monetary policy}

\begin{figure}[ht]
  \centering
  \includegraphics[height=8.2cm]{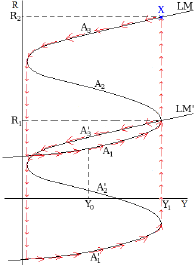}
  \caption{Cooperation between fiscal and monetary policy}
  \label{fig:cooperation_fiscal_monetary_policy}
\end{figure}

In this section we suggest how the cooperation between the fiscal and monetary policy can look like. The long-term real interest rate can quickly jump or fall as a consequence of the fiscal policy. The monetary policy can reduce these fluctuations by an appropriate decrease or increase of the expected inflation rate or of the constant part of the money stock. Our aim is an elimination of the negative influence of the fiscal policy which is the mentioned fluctuation of the long-term real interest rate. We demonstrate this cooperation on the model with once S-bent LM curve using Figure \ref{fig:cooperation_fiscal_monetary_policy}. We consider the case of the fiscal expansion, i.e. the increasing aggregate income. Let us start on the stable arc $A_1$ of the LM curve for the level $Y_0$. The government increases the aggregate income from $Y_0$ to $Y_1$, so the moving point goes along the stable arc $A_1$ to the point $[Y_1,R_1]$. In this point the moving point is attracted to the point $[Y_1,R_2]$, denoted by $X$. So, the long-term real interest rate is forced to jump to the level $R_2$ situated on the stable arc $A_3$. At this moment the central bank could increase the constant part of the money stock $M_S$ or the expected inflation rate $\pi^e$ to shift the LM curve downwards so that the point $X$ would become the point $[Y_1,R_1]$. The shifted LM curve is denoted by LM' on Figure \ref{fig:cooperation_fiscal_monetary_policy}. Now, the moving is still on the required stable arc (now denoted by $A'_3$) but the long-term real interest rate does not jump to the level $R_2$ and is still on the level $R_1$.

The central bank actions to reduce the falls of the long-term real interest rate are similar - at the right moment $M_S$ or $\pi^e$ should be decreased to shift the LM curve upwards.

\section{Illustrative example}

\begin{figure}[ht]
  \centering
  \includegraphics[height=6cm]{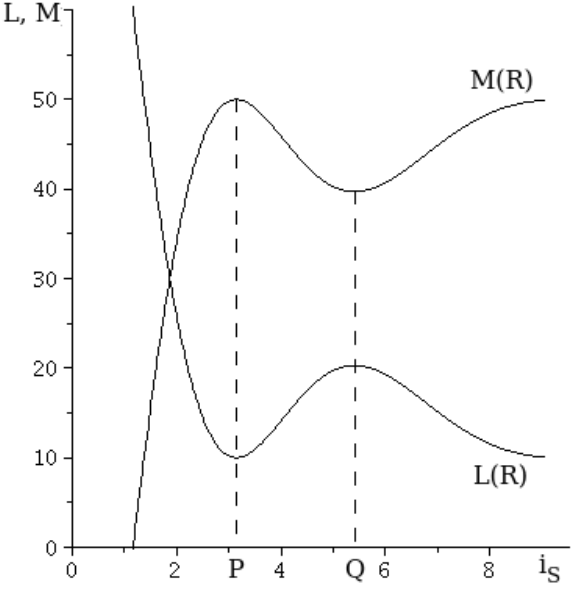}
  \caption{Graph of the functions $L(i_S)$ and $M(i_S)$ in the example}
  \label{fig:example_LR_MR}
\end{figure}

In this section, we demonstrate our modelling on the illustrative example. The demand for money and the supply of money function play the crucial role in our modelling, especially the demand for money and the supply of money function depending on the short-term nominal interest rate for some fixed aggregate income - $L(i_S)$ and $M(i_S)$. Thus firstly, we formulate $L(i_S)$ and $M(i_S)$ for some fixed $Y$ for observed interval $i_S \in [0,9]$: 
\begin{equation}
L(i_S)=-100 \cdot \frac{1-\text{e}^{-\frac{i_S}{2}} \cdot \cos(i_S)}{1+\text{e}^{-\frac{i_S}{2}}}+110,
\end{equation}
\begin{equation}
M(i_S)=100 \cdot \frac{1-\text{e}^{-\frac{i_S}{2}} \cdot \cos(i_S)}{1+\text{e}^{-\frac{i_S}{2}}}-50.
\end{equation}
The graphs of the functions $L(i_S)$ and $M(i_S)$ for suitable interval $i_S \in [0,9]$ and \linebreak $\text{P} \doteq 3.142$ and $\text{Q} \doteq 5.407$ are displayed on Figure \ref{fig:example_LR_MR}.
We select other dependencies as linear relationships because of a simplification. Then, the money demand function and the money supply function are given by the following formulas:
\begin{equation}
\label{example_L}
L(Y,R- MP + \pi^e)=-100 \cdot \frac{1-\text{e}^{-\frac{R+1}{2}} \cdot \cos(R+1)}{1+\text{e}^{-\frac{R+1}{2}}}+3Y+111 ,
\end{equation}
\begin{equation}
\label{example_M}
M(Y,R- MP + \pi^e)=100 \cdot \frac{1-\text{e}^{-\frac{R+1}{2}} \cdot \cos(R+1)}{1+\text{e}^{-\frac{R+1}{2}}}+2Y-50 ,
\end{equation}
where $L(Y)=3Y+1$, $M(Y)=2Y$ and $MP=0.5$, $\pi^e=1.5$ and thus $-MP+\pi^e=1$. The following formulas specify the investment and saving function:
\begin{equation}
\label{example_I}
I(Y,R)=\frac{Y}{5}-\frac{R}{6}+13 ,
\end{equation}
\begin{equation}
\label{example_S}
S(Y,R)=\frac{Y}{3}+\frac{R}{5}+8 .
\end{equation}
This investment function and this saving function fulfil all considered economic properties, i.e. (\ref{economic_I}), (\ref{economic_S}) and (\ref{economic_I_S}). For our observed interval the money demand function and the money supply function satisfy the properties (\ref{economic_L_R}), (\ref{economic_M_Y}) and (\ref{economic_M_R}) for $i_S \in [0,3.142) \cup (5.407,9]$ or $R \in [0,2.142) \cup (4.407,9]$, and the properties (\ref{economic_M_Y}) and (\ref{unusual_economic_L_M_R}) for $i_S \in (3.142, 5.407)$ or $R \in (2.142, 4.407)$. 

\begin{figure}[ht]
  \centering
  \includegraphics[height=6cm]{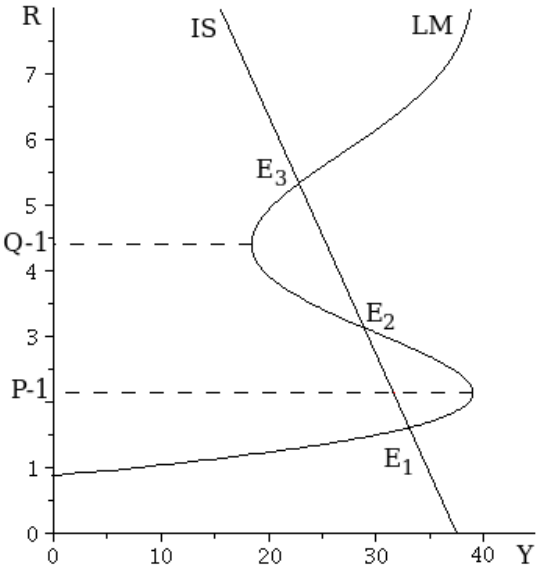}
  \caption{IS curve and LM curve in the example}
  \label{fig:example_IS_LM}
\end{figure}

If we consider $\alpha=2$ and $\beta=1.5$, then the new IS-LM model with the particular functions (\ref{example_L}), (\ref{example_M}), (\ref{example_I}) and (\ref{example_S}) is given by the system
\begin{equation}
\label{example_new_IS-LM_model}
\begin{array}{ll}
\textrm{IS:} & \frac{d Y}{d t} = -\frac{4}{15}Y-\frac{11}{15}R+10 \\
\textrm{LM:} & \frac{d R}{d t} = -300 \cdot \frac{1-\text{e}^{-{\frac{R+1}{2}}} \cdot cos(R+1)}{1+\text{e}^{-\frac{R+1}{2}}}+1.5Y+241.5  
\end{array}.
\end{equation}
The IS curve and the LM curve in this case are displayed on Figure \ref{fig:example_IS_LM}.
In this illustrative example the model (\ref{example_new_IS-LM_model}) has three singular points $E_1, E_2, E_3$, see Figure \ref{fig:example_IS_LM}, with approximate values:
\begin{displaymath}
\begin{array}{lll}
E_1: & Y_1 \doteq 33.077 , & R_1 \doteq 1.609 ; \\
E_2: & Y_2 \doteq 28.898 , & R_2 \doteq 3.128 ; \\
E_3: & Y_3 \doteq 22.834 , & R_3 \doteq 5.333 ; 
\end{array}  
\end{displaymath}
where the singular points $E_1$ and $E_3$ are the stable nodes and $E_2$ is the unstable saddle point. 

\begin{figure}[ht]
  \centering
  \includegraphics[height=6cm]{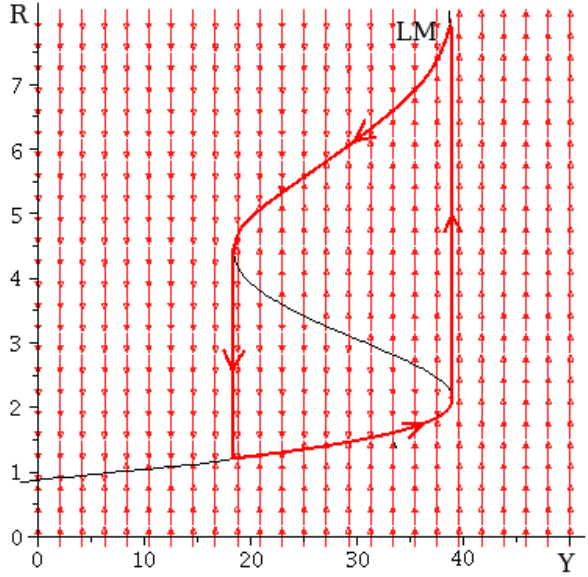}
  \caption{Relaxation oscillations in the example}
  \label{fig:example_rel_osc}
\end{figure}

The system (\ref{dynamic_model_LM_parameter_2}) with these particular functions is given by
\begin{equation}
\label{example_new_IS-LM_model_IS0}
\begin{array}{ll}
\textrm{IS:} & \frac{d Y}{d t} = 0 \\
\textrm{LM:} & \frac{d R}{d t} = -300 \cdot \frac{1-\text{e}^{-{\frac{R+1}{2}}} \cdot cos(R+1)}{1+\text{e}^{-\frac{R+1}{2}}}+1.5Y+241.5 
\end{array} .
\end{equation}
The sketch of the vector field given by the right-hand side of the system (\ref{example_new_IS-LM_model_IS0}) and the relaxation oscillations are depicted on Figure \ref{fig:example_rel_osc}. The cycle and the direction of the rotation is figured by the thick red line with arrows. 

\begin{figure}[ht]
  \centering
  \includegraphics[height=6cm]{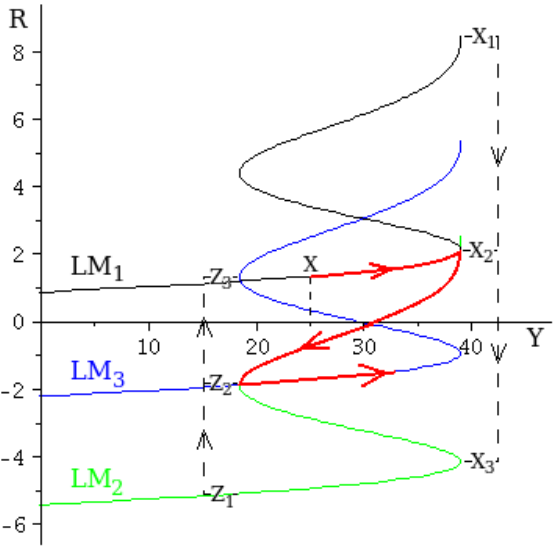}
  \caption{Cooperation between fiscal and monetary policy in the example}
  \label{fig:example_cooperation-fiscal_monetary_policy}
\end{figure}

Finally, we demonstrate the possible cooperation between the fiscal and monetary policy in this particular case. We consider that the changing of the aggregate income as \linebreak a consequence of the fiscal policy produces the relaxation oscillations on the money market, see Figure \ref{fig:example_rel_osc}. Then the LM curve is moved downwards or upwards as a consequence of the monetary policy, more precisely as a consequence of the changes of the expected inflation rate. We use Figure \ref{fig:example_cooperation-fiscal_monetary_policy} for the presentation of this cooperation.
Let us start at the black LM curve denoted by $\text{LM}_1$ in the point $X$. The fiscal expansion increases the aggregate income from the level $Y=25$ to the level $Y=39$ and the moving point goes very slowly from the point $X$ where $R \doteq 1.351$ to the point $X_2$ where $R \doteq 2.142$. In this point the moving point is attracted to the point $X_1$ and the long-term real interest rate is forced to quickly jump to the level $R \doteq 8.425$. At this moment the central bank increases an expected inflation rate from the level $\pi^e=1.5$ to the level $\pi^e \doteq 7.783$. As the consequence of this intervention the LM curve shifts downwards until the point $X_1$ becomes $X_2$. Now, the moving point is located in the point $X_2$ on the green LM curve denoted by $\text{LM}_2$ and the level of the long-term real interest rate is unchanged, i.e. $R \doteq 2.142$. Now, the moving point follows its standard path (but on the green LM curve denoted by $\text{LM}_2$) and goes very slowly from the point $X_2$ to the point $Z_2$ where $R \doteq -1.876$ and the $Y$ is slowly changed from the level $Y=39$ to the level $Y \doteq 18.411$. In $Z_2$ the moving point is attracted to the point $Z_1$ and the long-term real interest rate is forced to quickly fall down to the level $R \doteq -5.4019$. At this moment the central bank decreases an expected inflation rate from the level $\pi^e=7.783$ to the level $\pi^e \doteq 4.576$. As the consequence of this intervention the LM curve shifts upwards until the point $Z_1$ becomes $Z_2$. Then the moving point is located in the point $Z_2$ on the blue LM curve denoted by $\text{LM}_3$ and the level of the long-term real interest rate is unchanged, i.e. $R \doteq -1.876$. Now, the moving point follows its standard path (but on the blue LM curve denoted by $\text{LM}_3$) and goes very slowly from the point $Z_2$. Then $Y$ and $R$ very slowly increase as a consequence of the fiscal policy. And so on. The whole described path is displayed by the red line with arrows on Figure \ref{fig:example_cooperation-fiscal_monetary_policy}.

Similar mechanism can be applied for the second considered tool of the central bank, i.e. for the determination of the constant part of the money stock, or for the combination of these two tools.

\section*{Conclusion and discussion}

In these days, we encounter many fluctuations of different variables in the economies all of the world. The mainstream economics tries to explain these oscillations by standard clarifications but then another unexpected fluctuation will surprise us. In this paper we offer the alternative point of view. We provide some economic interpretation of our modelling but this explanation is newly created and does not succumb to the mainstream attitude. We created several models based on the new version of the IS-LM model and on the theory of the relaxation oscillations to explain the seemingly unexpected fluctuations of the long-term real interest rate. Then we pointed out the impacts of the fiscal and monetary policy on the economies described by these models. The fiscal policy can cause the relaxation oscillations on the money market and so the quick fluctuations of the long-term real interest rate and the monetary policy can reduced the forced falls or jumps of this variable.

\section*{Acknowledgements}
The research was supported by the Student Grant Competition of Silesian University in Opava, grant no. SGS/19/2010 and SGS/2/2013.

\end{document}